\documentstyle[11pt,epsf]{article}

\def\indspace{\hspace*{1.0em} }
\def\appendix{\setcounter{section}{0}
\def\thesection{Appendix \Alph{section}}
\def\theequation{\Alph{section}.\arabic{equation}}}
\newfont{\subsub}{cmr6}

\newcounter{szk}

\begin{document}
\title{Pareto law and Pareto index in the income distribution\\ of Japanese companies
}
\author{
\footnote{e-mail address: ishikawa@kanazawa-gu.ac.jp} Atushi Ishikawa
%$^a$ 
\\
%$^a$ 
Kanazawa Gakuin University, Kanazawa 920-1392, Japan
}
%\date{\today}
\date{}
\maketitle

%%%%%%%%%%%%%%%%%%%%%%%%%%%%%%%%%%%%%%%%%%%%%%
%%%       ABSTRACT
%%%%%%%%%%%%%%%%%%%%%%%%%%%%%%%%%%%%%%%%%%%%%%
\begin{abstract}
\indent
In order to study the phenomenon in detail that income distribution 
follows Pareto law, we analyze the database of high income companies 
in Japan. We find a quantitative relation between the average capital
of the companies and the Pareto index.
The larger the average capital 
becomes, the smaller the Pareto index becomes. From this relation, 
we can possibly explain that the Pareto index of 
company income distribution hardly changes, while the Pareto index 
of personal income distribution changes sharply, from a viewpoint 
of capital (or means). We also find a quantitative relation 
between the lower bound of capital and the typical scale at which 
Pareto law breaks.
The larger the lower bound of capital becomes,
the larger the typical scale becomes. From this result, the reason 
there is a (no) typical scale at which Pareto law breaks in the 
income distribution can be understood through (no) constraint, 
such as the lower bound of capital or means of companies, 
in the financial system.
\end{abstract}

\begin{flushleft}
PACS code : 04.60.Nc\\
Keywords : Econophysics; Income of companies;
Pareto law; Pareto index; Typical scale
\end{flushleft}

\vspace{1cm}
%%%%%%%%%%%%%%%%%%%%%%%%%%%%%%%%%%%%%%%%%%%%%%
%%%       SECTION
%%%       INTRODUCTION
%%%%%%%%%%%%%%%%%%%%%%%%%%%%%%%%%%%%%%%%%%%%%%
\section{Introduction}
\label{sec-introduction}
\indspace
It is well known that personal high income distribution follows 
power law \cite{Pareto}:
\begin{eqnarray}
    N_P(\ge x) \propto x^{-\alpha}~
    \label{Pareto}
\end{eqnarray}
with high accuracy \cite{ASNOTT}.
Here $x$ is the income and $N_P(\ge x)$ is the number of persons
with the income larger than or equals to $x$,
and the power law and the exponent $\alpha$ is called 
Pareto law and Pareto index, respectively.

For instance in Japan, persons who paid annual income tax more than 10 million yen
are announced publicly as ``high income taxpayers'' every year.
It is reported that their income distribution
follows Pareto law \cite{ASNOTT}.
In Ref.~\cite{FSAKA},
it is also reported that Pareto index annually changes around $2.0$,
and especially changed sharply ($1.8 \Rightarrow 2.1$) when the economic bubble collapsed
in Japan.

In addition, it is well known that,
not only the personal high income distribution,
but also the company high income distribution follows Pareto law.
For instance in Japan, companies having annual income more than 40 million yen
are announced publicly as ``high income companies'' every year.
It is reported that their income distribution follows Pareto law
and the Pareto index hardly changes around $1.0$ annually
\cite{OTT, Mizuno, FGAGS}.
Power law with exponent $-1$ is especially called Pareto-Zipf law. 
Furthermore in Ref.~\cite{OTT, Mizuno},
it is reported that income distributions in most job categories
follow Pareto law,
however,
those Pareto indices scatter around $1.0$.

On the other hand,
it is well known that personal or company
distribution with low-middle income region does not follow Pareto law
\cite{Gibrat, Badger, ASNOTT, AIST}.

The research on income distribution is well investigated 
in econophysics \cite{MS},
and many models to explain the income distribution are proposed
(For instance \cite{FSAKA, Mizuno, FGAGS, AIST}).
There are, however, several unsolved problems. 
The first biggest question is why Pareto law appears (breaks). 
We want to clarify the reason
that the high income distribution follows Pareto law,
while the low-middle income distribution does not follow Pareto law.
The second question is why Pareto index changes (does not change). 
For instance in recent annual income distribution in Japan,
we want to understand the reason
that Pareto index of personal high income distribution
changes sharply,
while Pareto index of company high income distribution hardly changes.
The third question is whether
there is any economical variable related to Pareto index.
If we find an answer to the third question,
we may be able to solve the second and the first questions.

In order to clear these problems,
we should examine Pareto law in detail.
In this paper, we analyze the database of high income companies in recent Japan. 
The reason is as follows.
The information in the database of high income taxpayers is limited,
only income tax, individual name and the address.
The database of high income companies, however, includes rich financial information,
not only income, company name and the address,
but also capital, number of employees, sales and profits.
We analyze the database published by TOKYO SHOKO RESEARCH, LTD. \cite{TSR}.

By analyzing this database,
we obtain the following several relations between Pareto law and capital.
Firstly, by classifying companies into job categories,
we find a quantitative relation between the average capital
of companies in each job category and the Pareto index.
The larger the average capital becomes, the smaller the Pareto index becomes.

Secondly, we classify companies into capital categories.
We observe that
even high income distribution 
follows power law only in the large-scale region
and deviates further from the power law in the smaller-scale region.
This distribution has the same feature
to the distribution with low-middle income region.
In this capital classification,
we also observe that the larger the
capital class
becomes, the smaller the Pareto index in the power law region becomes.
This means that the relation between the average capital and the Pareto index,
found by classifying companies into job categories,
holds regardless of the job classification. 
In addition,
we observe that the larger the
capital class becomes, the larger the
typical scale at which Pareto law breaks becomes.
In this paper, we call the scale at which Pareto law breaks
the typical scale \cite{AIST}.

Thirdly, we classify companies into cumulative capital categories.
We observe that, in this classification,
each Pareto index in power law region can be estimated to be about $1.0$.
When Pareto index is fixed in each class, the
typical scale can be read systematically. 
As a result, we find a quantitative relation 
between the lower bound of cumulative capital category and the typical scale.
The larger the lower bound of capital becomes, the larger the 
typical scale becomes.

From these analytic results, we obtain the following conclusions. 
Firstly, the average capital is related to the Pareto index.
From this relation,
the small change of Pareto index can be interpreted  
as small fluctuations in the average capital. 
Secondly, the typical scale is related to the lower bound of capital.
As a result,
the reason Pareto law appears in the income distribution
can be understood through no constraint, 
such as the lower bound of capital (or means), 
in the financial system \cite{AIST}.
%%%%%%%%%%%%%%%%%%%%%%%%%%%%%%%%%%%%%%%%%%%%%%%%%%
%%%       SECTION
%%%       Income distribution in job categories
%%%%%%%%%%%%%%%%%%%%%%%%%%%%%%%%%%%%%%%%%%%%%%%%%%
\section{Income distribution in job categories}
\label{sec-category}
\indspace
In this Section, we classify Japanese high income companies
into job categories, and
explain that the average capital
of the companies in each job category is related to the Pareto index.
Before the analysis, we describe our database.

In Japan, companies having annual income more than 40 million yen
are announced publicly as ``high income companies'' every year.
For example, we can obtain the lists of all the companies
in the database published by DIAMOND INC. 
which is supplied from TEIKOKU DATABANK, LTD. (TDB).
The information for each company in the database is, however, limited, 
only company's name, its job category and the income.
In this paper, we analyze the database, ``CD Eyes'' published by 
TOKYO SHOKO RESEARCH, LTD. (TSR) \cite{TSR} instead of the TDB database.
The TSR database includes rich financial information,
not only company's name, its job category and the income,
but also capital, number of employees, sales and profits.
We should note that
these detailed pieces of information are not added to all the high income companies announced publicly. 
The number of the high income
companies to which the detailed information is added in this database
is smaller than the number of the high income companies announced publicly. 
However, we consider that there is a sufficient amount of data in this database
to investigate Pareto law and Pareto index. 

\begin{figure}[htb]
 \centerline{\epsfxsize=0.95\textwidth\epsfbox{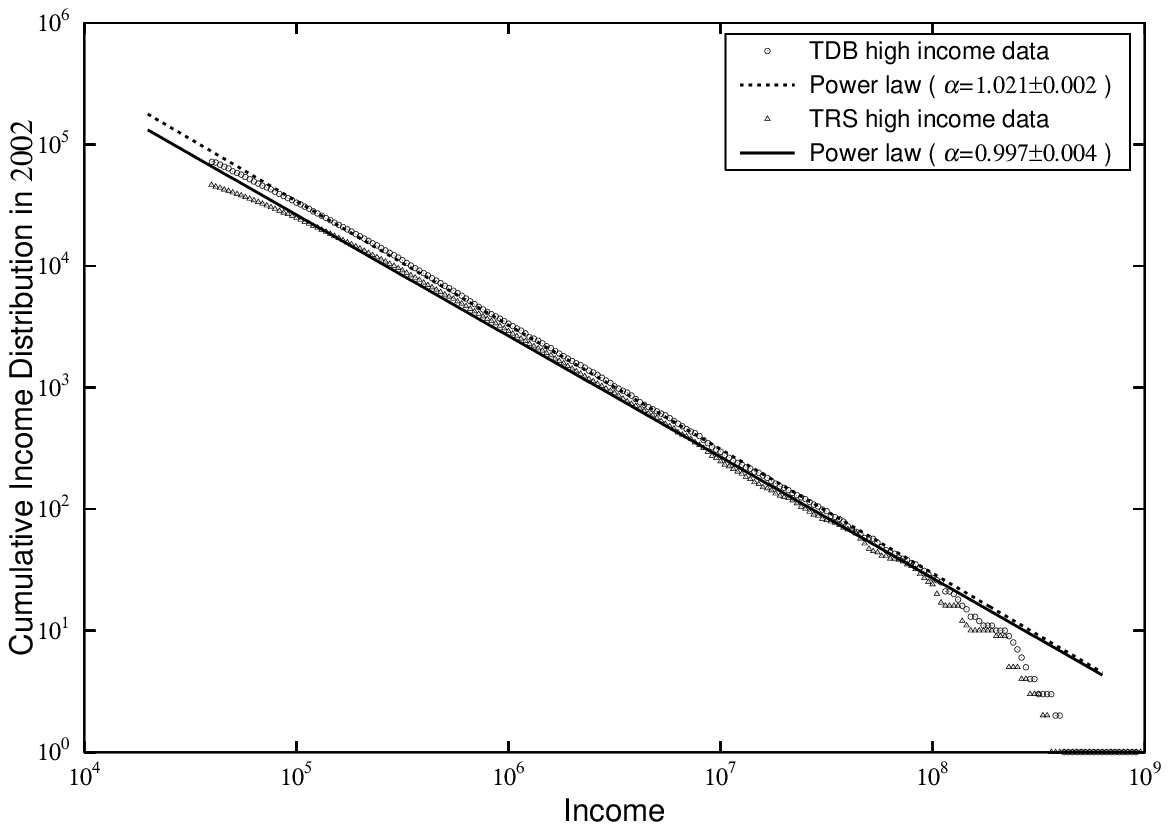}}
 \caption{Company income distributions in the fiscal year 2002 Japan.}
 \label{TDBandTSR}
%\end{figure}
%%%%%%%%%%%%%%%%%%%%%%%%%%%%%%%%%%%%%%%%%%%%%%%%%%%%%%%%%
%%%%%%%%%%%%%%%%%%%%%%%%%%%%%%%%%%%%%%%%%%%%%%%%%%%%%%%%%
%\begin{figure}[htb]
 \centerline{\epsfxsize=0.95\textwidth\epsfbox{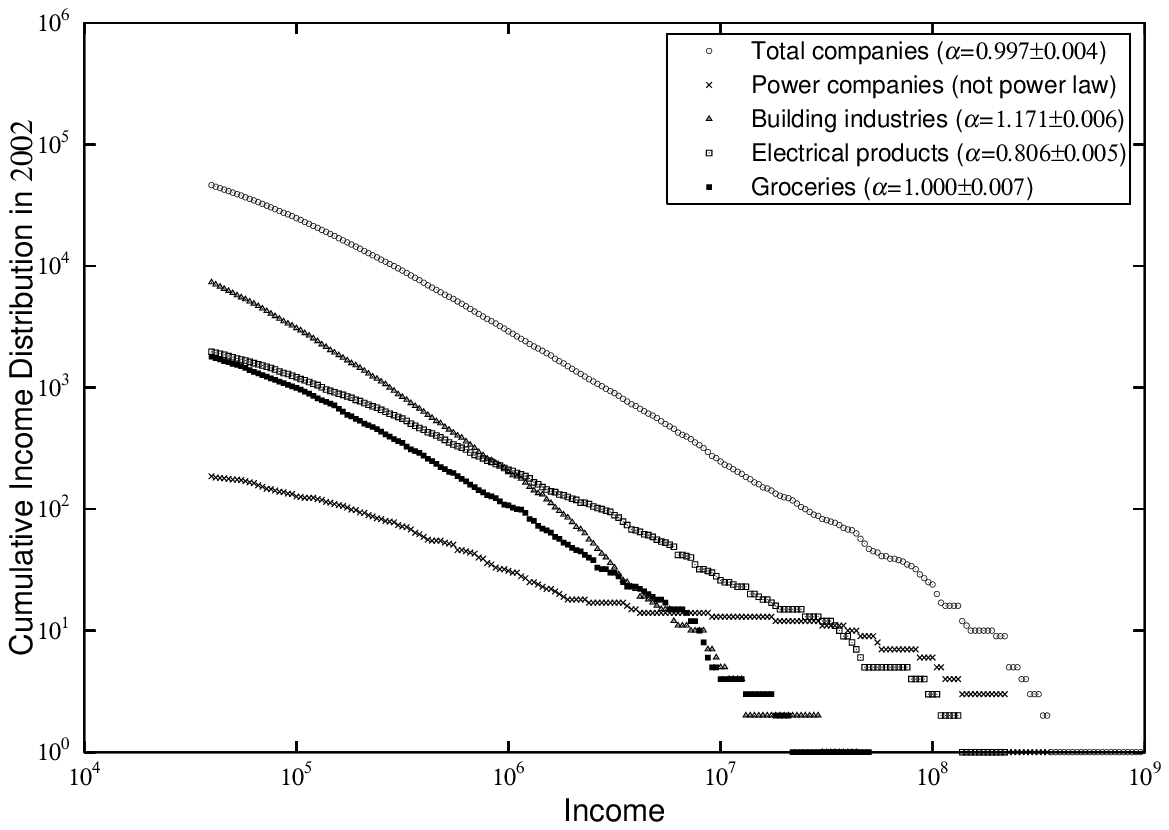}}
 \caption{Company income distributions in typical job categories.}
 \label{TSR}
\end{figure}

For example,
we compare two income distributions of Japan in the fiscal year 2002 (Fig.~\ref{TDBandTSR}).
One is obtained from the TDB data which includes all the high income
companies announced publicly,
the other is obtained from the TSR data which does not include all the companies.
The horizontal axis indicates the income $x$ in units of thousand yen
and the vertical axis indicates the number of companies 
with the income larger than or equals to $x$ ($N_P(\ge x)$ in Eq.~(\ref{Pareto})).
There is almost no difference in two distributions 
except for a parallel gap along the vertical axis.
Moreover, if we estimate Pareto index up to 2 figures of significant figures, 
the difference between both Pareto indices can be disregarded. 
Here, we estimate the Pareto index for each data
by excluding top $0.1\%$ and bottom $10\%$ and using least-square-fit.

We also show several income distributions in typical job categories for the TRS data
in Fig.~\ref{TSR}.
Similar distributions are also obtained from the TDB data \cite{OTT, Mizuno}.
We estimate the Pareto index of the distribution in each job category by
excluding top $1\%$ and bottom $10\%$.
Therefore, we consider that the TSR data
is useful to investigate Pareto law and Pareto index. 

We estimate Pareto index of power law distribution in every job category using the TSR data
in the fiscal year 2002. 
There are exceptions, for example, 
the distribution for power companies in Fig.~\ref{TSR}
should be considered to deviate from power law.
Such deviations from power law can be measured
by coefficient of determination in least-square-fit.
In the TSR database, we can know the capital of each company,
therefore, we also calculate the average capital of companies 
in each job category ($\bar{C}$).
We show the relation between the average capital
and the Pareto index (Fig.~\ref{Capital_vs_ParetoIndex}).
The horizontal axis indicates 
the average capital in units of thousand yen
and the vertical axis indicates the Pareto index ($\alpha$).
Here,
we exclude several job categories in which
income distribution does not follow power law.
We decide the exclusion by the criterion that
the coefficient of determination is less than $0.97$.
In Fig.~\ref{Capital_vs_ParetoIndex},
we observe that the larger the average capital of each job category becomes,
the smaller the Pareto index becomes,
and we quantitatively find
\begin{eqnarray}
    \alpha \simeq - 0.31 \log \bar{C} + 2.7~.
    \label{CapitalvsParetoIndex}
\end{eqnarray}

Is this relation peculiar to the job classification?
In order to examine this question, in the next Section,
we classify companies into capital categories regardless of the job category.
%%%%%%%%%%%%%%%%%%%%%%%%%%%%%%%%%%%%%%%%%%%%%%%%%
%%%       SECTION
%%%       Income distribution in capital categories
%%%%%%%%%%%%%%%%%%%%%%%%%%%%%%%%%%%%%%%%%%%%%%%%%
\section{Income distribution in capital categories}
\label{sec-class}
\begin{figure}[htb]
 \centerline{\epsfxsize=0.95\textwidth\epsfbox{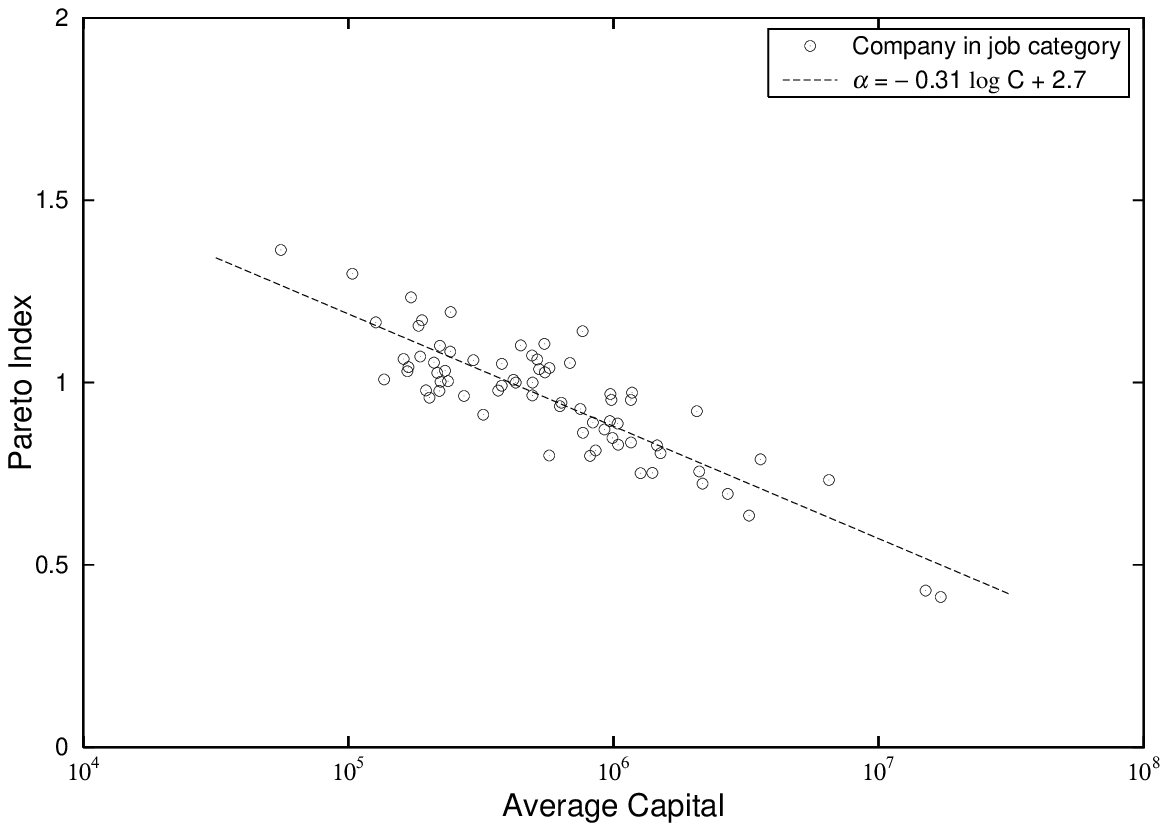}}
 \caption{Relation between the average capital of companies and the Pareto index.}
 \label{Capital_vs_ParetoIndex}
%\end{figure}
%%%%%%%%%%%%%%%%%%%%%%%%%%%%%%%%%%%%%%%%%%%%%%%%%%%%%%%%%
%%%%%%%%%%%%%%%%%%%%%%%%%%%%%%%%%%%%%%%%%%%%%%%%%%%%%%%%%
%\begin{figure}[htb]
 \centerline{\epsfxsize=0.95\textwidth\epsfbox{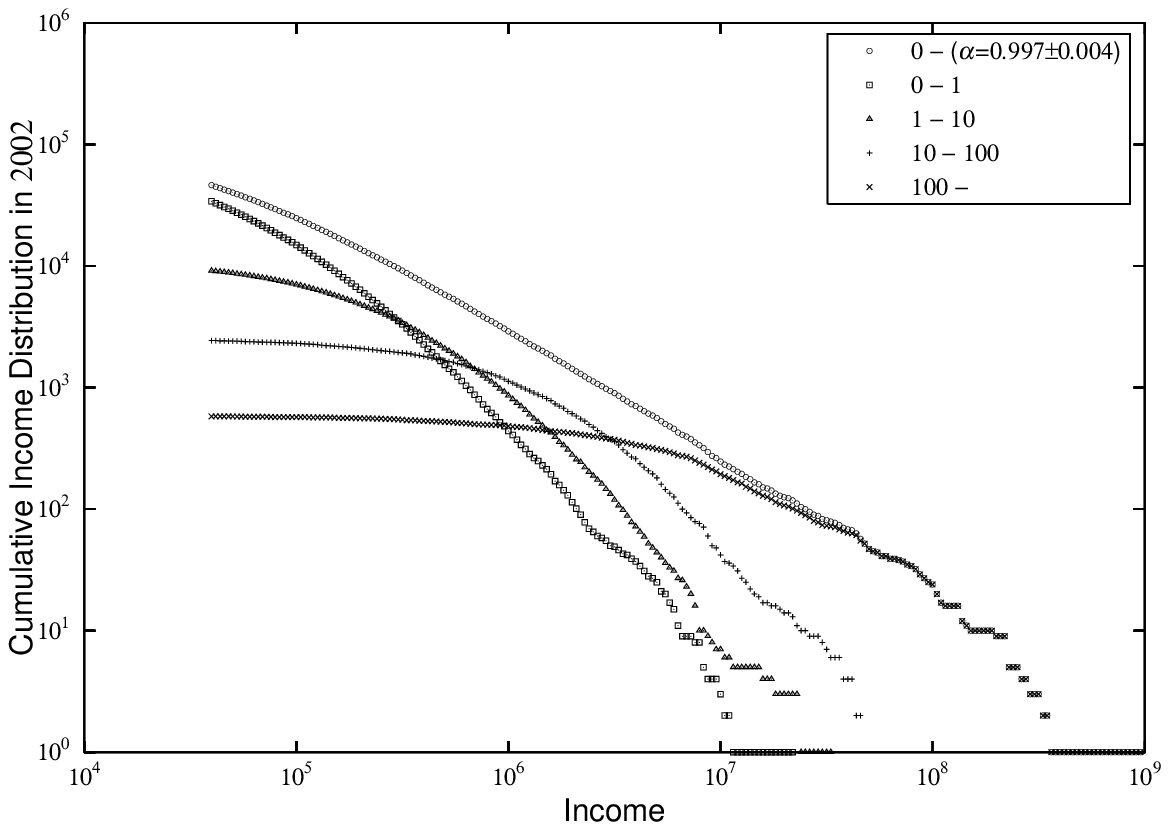}}
 \caption{Company income distributions in capital categories.}
 \label{TSRCapitalClass}
\end{figure}

\indspace
In this Section, we classify Japanese high income companies
into capital categories such as
$0 - 1$, $1 - 10$, $10 - 100$ and $100 -$
in units of 100 million yen.
In this classification, we find that
even high income distribution has not only the
power law region but also the region which does not follow the power law
(Fig.~\ref{TSRCapitalClass}).
In Fig.~\ref{TSRCapitalClass},
we can observe that the larger capital class becomes, 
the smaller the Pareto index in power law region becomes.
This qualitative relation is similar to Eq.~(\ref{CapitalvsParetoIndex})
obtained in the job classification.
Consequently, we consider that
the relation between the average capital and the Pareto index (\ref{CapitalvsParetoIndex})
is directly nothing to the job classification.
We also observe that 
the larger capital class becomes,
the larger the
typical scale becomes in Fig.~\ref{TSRCapitalClass}.

In Fig.~\ref{TSRCapitalClass}, however,
it is difficult to read Pareto index and typical scale
quantitatively,
because it is hard to determine the power law region and
the typical scale.
In order to avoid this difficulty,
in the next Section,
we classify companies into cumulative capital categories.
%%%%%%%%%%%%%%%%%%%%%%%%%%%%%%%%%%%%%%%%%%%%%%%%%%
%%%       SECTION
%%%       Income distribution in cumulative capital categories
%%%%%%%%%%%%%%%%%%%%%%%%%%%%%%%%%%%%%%%%%%%%%%%%%%
\section{Income distribution in cumulative capital categories}
\label{sec-cumclass}
\begin{figure}[htb]
 \centerline{\epsfxsize=0.95\textwidth\epsfbox{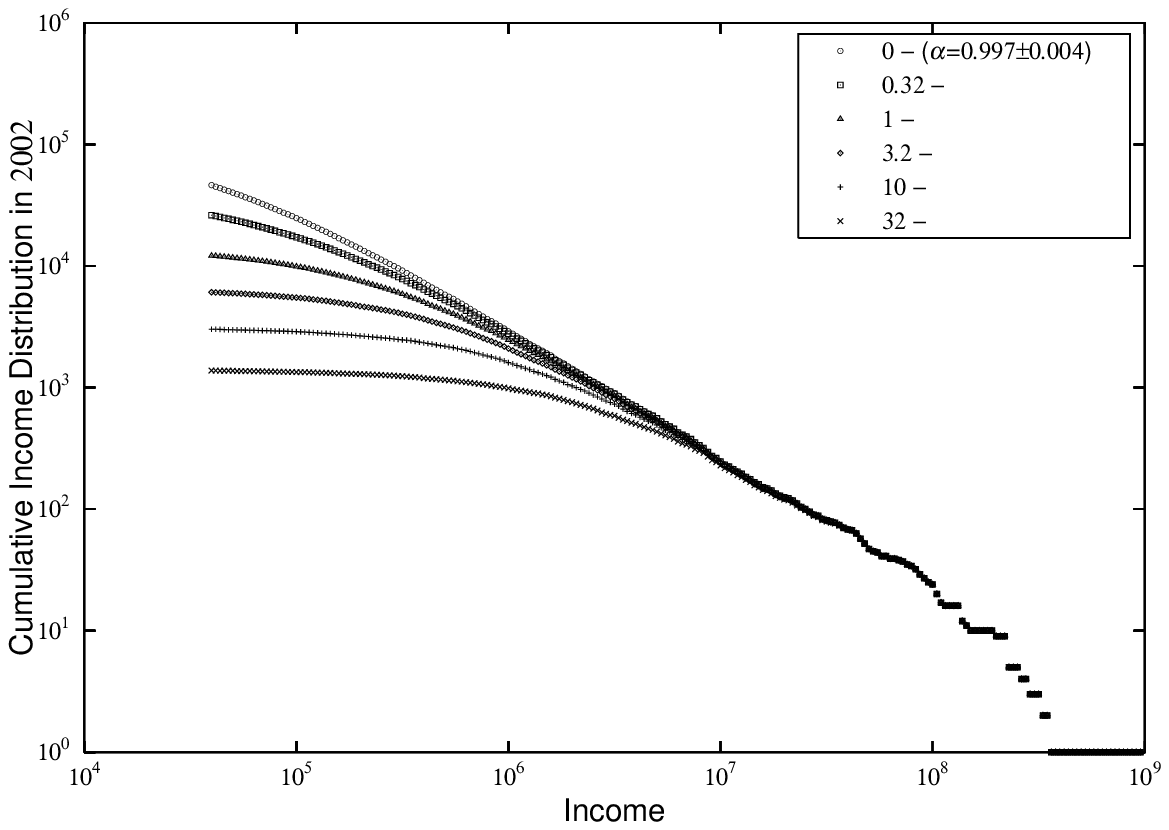}}
 \caption{Company income distributions in cumulative capital categories.}
 \label{TSRCapitalCumClass}
%\end{figure}
%%%%%%%%%%%%%%%%%%%%%%%%%%%%%%%%%%%%%%%%%%%%%%%%%%%%%%%%%
%%%%%%%%%%%%%%%%%%%%%%%%%%%%%%%%%%%%%%%%%%%%%%%%%%%%%%%%%
%\begin{figure}[htb]
 \centerline{\epsfxsize=0.95\textwidth\epsfbox{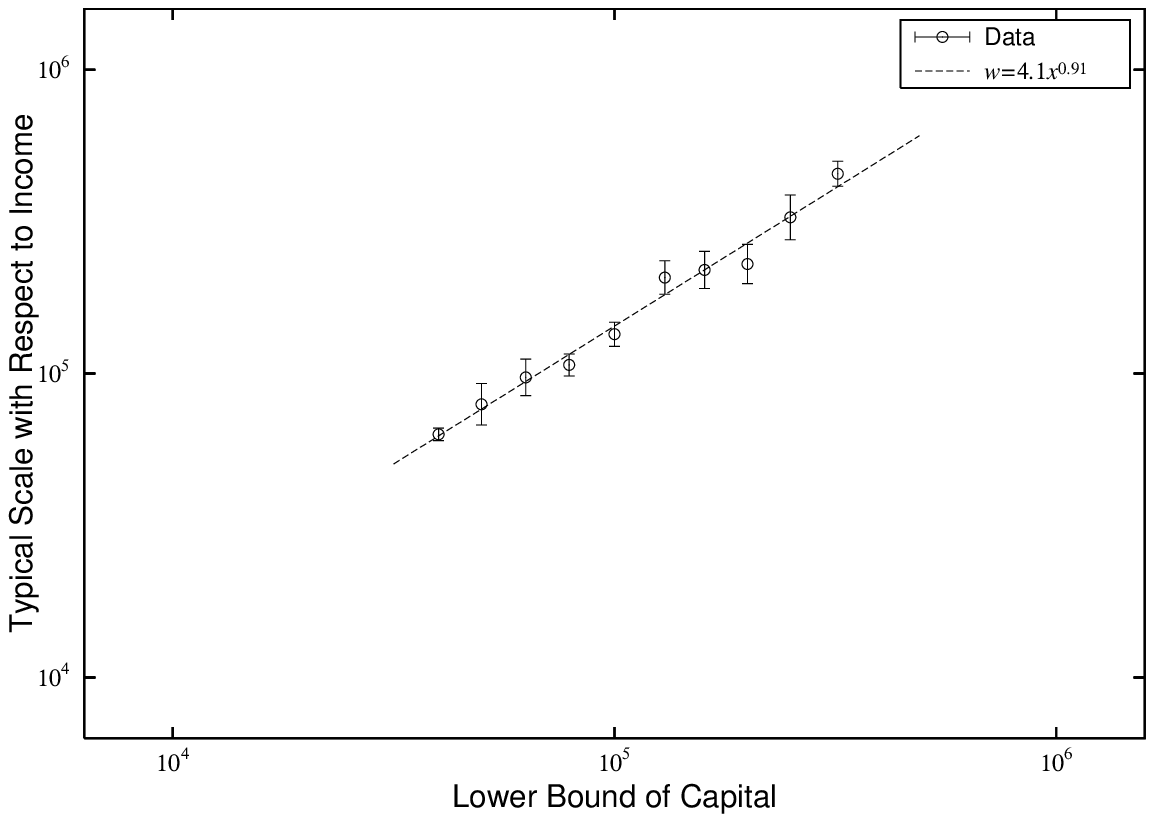}}
 \caption{Relation between the lower bound of capital and the typical scale.}
 \label{TSRTipicalScale_vs_Capital}
\end{figure}

\indspace
From the distributions in Fig.~\ref{TSRCapitalClass},
we suppose that
companies in the large capital class leads Pareto-Zips law
in the large-scale region.
In order to confirm this,
we classify companies into cumulative capital categories such as
$0-$, $0.32-$, $1-$, $3.2-$, $10-$ and $32-$
in units of 100 million yen.
In this classification, we find that
each Pareto index in power law region
can be estimated to be about $1.0$
(Fig.~\ref{TSRCapitalCumClass}).

When Pareto index is fixed in each class, 
the typical scale can be read systematically. 
We read the scale
by using the distribution functions \cite{ITY, AIST}
derived in 2-dim $R^2$ gravity model \cite{KN}
(See Appendix in detail).
As a result,
we observe that the larger the lower bound of capital in each cumulative capital category ($C_{min}$)
becomes, the larger
the typical scale ($w$) becomes (Fig.~\ref{TSRTipicalScale_vs_Capital}).
In Fig.~\ref{TSRTipicalScale_vs_Capital},
we find a quantitative relation:
\begin{eqnarray}
    w \simeq 4.1~{C_{min}}^{0.91}~.
    \label{Cmin_vs_TipicalScale}
\end{eqnarray}

In this relation,
when there is no bound with respect to capital ($C_{min}=0$),
there is no typical scale ($w=0$) and
the income distribution follows Pareto law in all scale region.
Eq.~(\ref{Cmin_vs_TipicalScale}) is, therefore,
consistent with the observation
that almost all distributions follow
Pareto law in the job classification in Section \ref{sec-category}.
%%%%%%%%%%%%%%%%%%%%%%%%%%%%%%%%%%%%%%%%%%%%%%
%%%       SECTION
%%%       DISCUSSION
%%%%%%%%%%%%%%%%%%%%%%%%%%%%%%%%%%%%%%%%%%%%%%
\section{Discussion}
\label{sec-discussion}
\indspace
In this paper,
by analyzing the database 
of high income companies
in the fiscal year 2002 Japan,
we obtained the following two results.
Average capital is related to the Pareto index (Eq.~(\ref{CapitalvsParetoIndex})).
Typical scale at which Pareto law breaks is related to the lower bound of capital
(Eq.~(\ref{Cmin_vs_TipicalScale})).

Firstly, from Eq.~(\ref{CapitalvsParetoIndex}), the
small change of Pareto index can be interpreted  
as small fluctuations in the average capital. 
If we assume that similar equation to Eq.~(\ref{CapitalvsParetoIndex}) holds
in other years, we may explain that the Pareto index of company income distribution hardly changes.
In addition,
if we assume that capital and assets have a relation
and similar equation to Eq.~(\ref{CapitalvsParetoIndex}) holds in personal income distribution,
we may also explain that the Pareto index of personal income distribution changes sharply
by large fluctuations in average means.

Secondly, from Eq.~(\ref{Cmin_vs_TipicalScale}),
the reason 
there is a (no) typical scale at which Pareto law breaks down in the income distribution
can be understood through (no) constraint, 
such as the lower bound of capital or means, 
in the financial system \cite{AIST}.

There is an unsolved problem:
why the exponent of the power law distribution of all high income companies is $1$.
It is reported that the
Pareto index in Italy, Spain, France and UK also hardly changes around $1$ \cite{FGAGS}.
It is natural to consider that
there is some mechanism which fixes Pareto index at $1$ universally.
The relation between the average capital and the Pareto index (\ref{CapitalvsParetoIndex})
may be used to study this problem.
We should, however, determine the normalization
of Eq.~(\ref{CapitalvsParetoIndex}) to study this problem.
Although Pareto index $\alpha$ in the left side of Eq.~(\ref{CapitalvsParetoIndex})
is dimensionless, the
average capital $\bar{C}$ in the right side has monetary dimension.
Since logarithm of the average capital is taken,
Eq.~(\ref{CapitalvsParetoIndex}) is sufficient to estimate change of the Pareto index
with respect to the average capital.
When we estimate magnitude of the Pareto index, however,
we must divide the average capital by some criterion with monetary dimension
and determine the normalization.
It is difficult to fix this normalization
only by analyzing a single fiscal year.

We should examine the relation between the average capital and the Pareto index over several years
to complete Eq.~(\ref{CapitalvsParetoIndex}) up to the normalization.
This remains to study.

\newpage
%%%%%%%%%%%%%%%%%%%%%%%%%%%%%%%%%%%%%%%%%%%%%%
%%%%%%%%%%%%%%%%%%%%%%%%%%%%%%%%%%%%%%%%%%%%%%

\section*{Acknowledgements}
\indent

The author would like to thank
H.~Itoh and TOKYO SHOKO RESEARCH, LTD.
for kindly providing high-quality data.
He is also grateful to the Yukawa Institute for Theoretical 
Physics at Kyoto University,
where this work was initiated during the YITP-W-03-03 on
``Econophysics - Physics-based approach to Economic and
Social phenomena -'',
and specially to Professor.~H. Aoyama for useful comments.
Thanks are also due to Professor.~H. Terao and Professor.~H. Kasama
for careful reading of the manuscript.

%%%%%%%%%%%%%%%%%%%%%%%%%%%%%%%%%%%%%%%%%%%%%%%%%%%%%%%%%%%%%%%%%%%%%%%%%%%%%%
%%%%%%%%%%%%%%%%%%%%%%%%%%%%%  Appendix  %%%%%%%%%%%%%%%%%%%%%%%%%%%%%%%%%%%%%
%%%%%%%%%%%%%%%%%%%%%%%%%%%%%%%%%%%%%%%%%%%%%%%%%%%%%%%%%%%%%%%%%%%%%%%%%%%%%%
\appendix
\section{2-dim $R^2$ gravity model}
\indspace
In Section~\ref{sec-cumclass},
we used the distribution functions \cite{ITY, AIST}:
\begin{eqnarray}
   n_W(x)&\sim& N_W~x^{\gamma_0-2}~
   \exp\biggl[ -\frac{w}{x}
   \biggr]
   ~~~~~~   {\rm for}~~~~ 0 <  x < w~,
   \label{Weibull Aproximation}\\
   n_P(x)&\sim&  N_P~x^{\gamma_{\infty}-2}   ~~~~~~~~~~~~~~~~~~~~
   {\rm for}~~~~ w < x < \infty
   \label{Pareto Aproximation}
\end{eqnarray}
derived in 2-dim $R^2$ gravity model \cite{KN}
in order to read the typical scale at which power law breaks down. 
Here $\gamma_0$ and $\gamma_{\infty }$ are constants
determined by the parameter $c$,
\begin{eqnarray}
    \gamma_0&=&2+\frac{c-12}{6}~,
    \label{g0}
    \\
    \gamma_{\infty }&=&2+\frac{c-25-\sqrt{(25-c)(1-c)}}{12}~.
    %+\frac{d}{24 \pi}m^2 A.
    \label{ginfty}                             
\end{eqnarray}

In the power law region of the company income distribution,
we observed that 
the exponent $\gamma_{\infty }-2$ of power law 
(\ref{Pareto Aproximation}) is $-2$,
%of power law function (\ref{Pareto Aproximation})
then we decided that $c=1$ and $\gamma_0-2=-11/6$.
We should note that the exponent of power law
is $-1$
when the number of companies is accumulated,
$N_P(\ge x)=\int^{\infty}_x dx~n_p(x) \propto x^{-1}$.
The typical scale is $w$,
which connects two distribution functions 
(\ref{Pareto Aproximation}) and (\ref{Weibull Aproximation}).
We read the scale $w$ by fitting the distribution function (\ref{Weibull Aproximation}) 
to the region where 
the distribution does not follow power law \cite{AIST}.

The first reason we use 2-dim $R^2$ gravity model,
it can derive both power law distribution and distribution which does not follow the power law
in a unified framework.
The typical scale at which power law breaks, therefore, is clearly understood in this model.
We introduce a scale $w$ into the model, which is originally scale free $S_{free}$,
by adding an interaction term with a scale $S_{scale}$ \cite{KN}:
\begin{eqnarray}
    S_{total}&=&S_{free}+S_{scale}~,\\
    \label{action total}
    S_{free}(X^i;g)&=&\frac{1}{8 \pi} \int {\rm d^2}x \sqrt{g} g^{\mu \nu} 
    \partial_{\mu} X^i \partial_{\nu} X_i~,
    \label{X}\\
    S_{scale}(g)&=&\frac{w}{64 \pi^2} \int {\rm d^2}x \sqrt{g} R^2~.
    \label{R^2}
\end{eqnarray}
Here, $S_{total}$ is the total action,
$X^i(i=1,2,\cdots ,c)$ are conformal scalar matter fields coupled with gravity,
$g_{\mu \nu}(\mu, \nu=0, 1)$ is the metric of 2-dim surface,
$R$ is the scalar curvature
and $w$ is a coupling constant of length dimension $2$.
The partition function for fixed area $A$ of 2-dim surface is
\begin{eqnarray}
    Z(A)=\int \frac{{\cal D} g 
    {\cal D} X}{\rm vol(Diff)}
    {\rm e}^{-S_{total}(X^i;g)}~ \delta(\int d^2 x \sqrt{g}-A)~.
    \label{partition}
\end{eqnarray}
By using this partition function,
the statistical average number of
finding a surface of area $B$ on a closed surface of area $A$
can be expressed as 
Eqs.~(\ref{Weibull Aproximation}) and (\ref{Pareto Aproximation}) \cite{KN, JM, ITY, AIST}.
Here we used expression $(1-B/A)B=x$.

\begin{figure}[htb]
 \centerline{\epsfxsize=0.65\textwidth\epsfbox{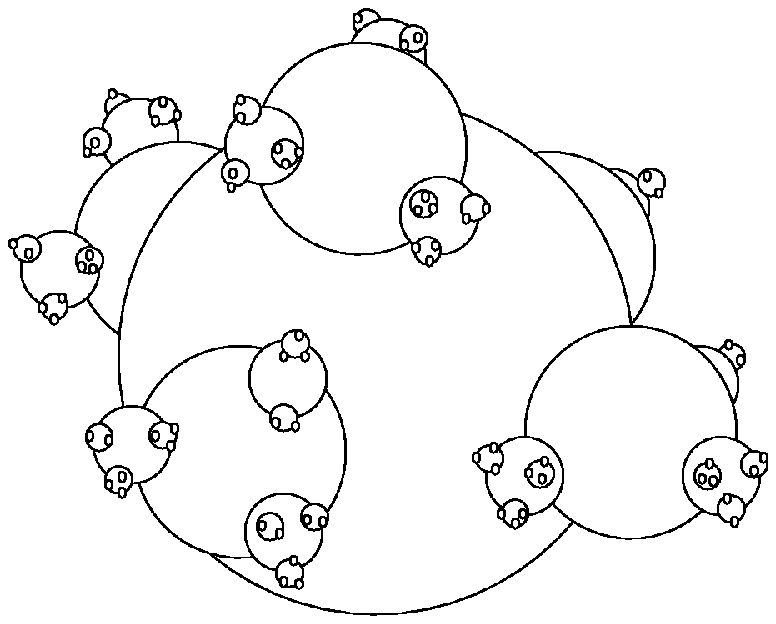}}
 \caption{A typical 2-dim surface is fractal.}
 \label{2-dim Random Surface}
%\end{figure}
%%%%%%%%%%%%%%%%%%%%%%%%%%%%%%%%%%%%%%%%%%%%%%%%%%%%%%%%%
%%%%%%%%%%%%%%%%%%%%%%%%%%%%%%%%%%%%%%%%%%%%%%%%%%%%%%%%%
%\begin{figure}[htb]
 \centerline{\epsfxsize=0.95\textwidth\epsfbox{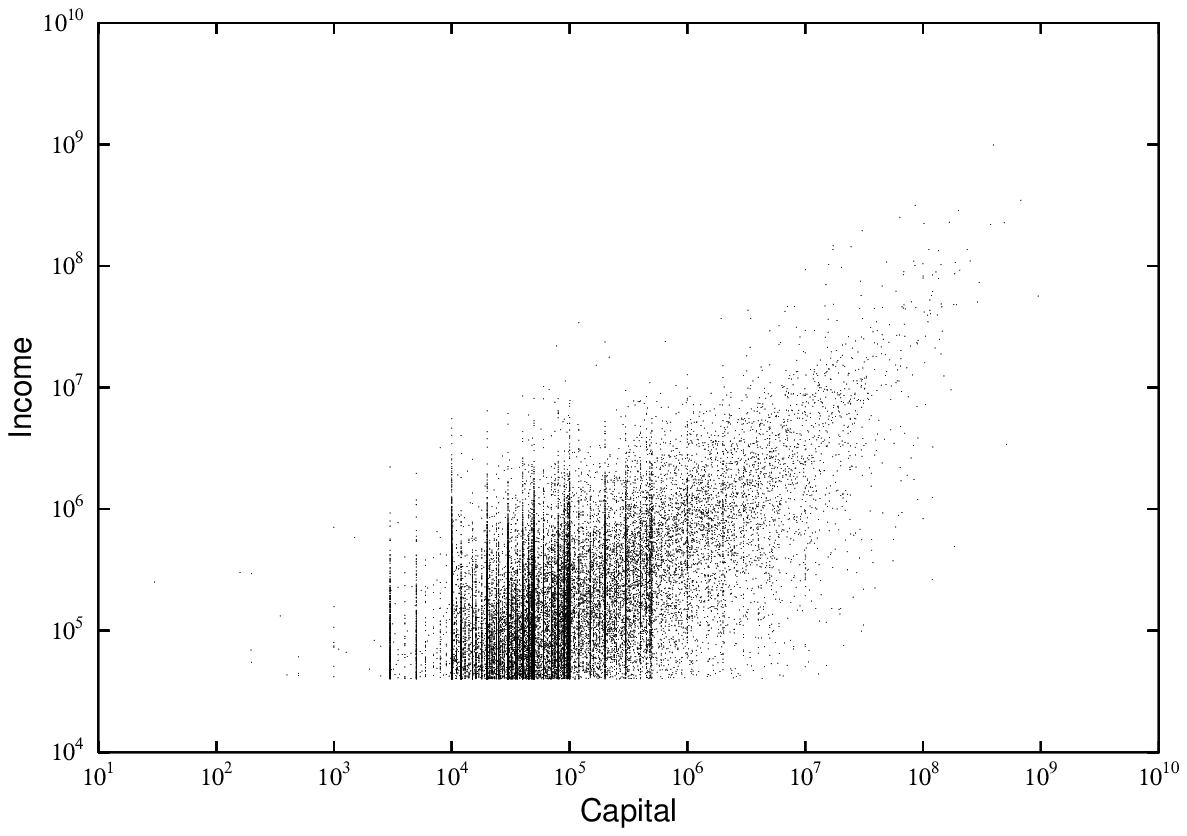}}
 \caption{Relation between the capital and the income of high income companies.}
 \label{Income}
\end{figure}

The distribution functions (\ref{Weibull Aproximation}) and (\ref{Pareto Aproximation})
are geometrically understood as follows.
The scale free action $S_{free}$ leads
the locality that local structure of 2-dim surface does not 
affect the 2-dim structure away from that place. From this property,
it is well known that a typical 2-dim surface has fractal self-similar structure
(Fig.~\ref{2-dim Random Surface}) \cite{BIPZ, KAD}.
On the other hand,
the scale variant action $S_{scale}$ has the effect to
let 2-dim surface flat.
When this effect cannot be ignored, 
local structure of 2-dim surface does
affect the 2-dim structure away from that place.
The locality of 2-dim surface breaks down and
some kind of constraint is imposed on 2-dim surface.
Therefore,
the statistical average number of
finding a small surface of area $B$ on a closed surface of area $A$
decreases.
As a result, 
the fractal structure in the small area region breaks down
and the typical scale is introduced into the system
\cite{KN, ITY}. 

This 2-dim geometric phenomenon has qualitative common features
to the phenomenon in the company income distribution. 
When there is lower bound of capital,
the company income whose capital is below the lower bound
is hardly counted in the company income distribution (Fig.~\ref{Income}).
As a result, 
the small income distribution does not follow Pareto law
and the scale at which fractal power law breaks is introduced into the income distribution.
From this qualitative similarity between two kinds of distributions, 
we used the distribution functions derived in 2-dim $R^2$ gravity model
as measuring tool to read the typical scale in the income distribution \cite{AIST}.

%------------------ Bibliography -------------------------

%--------------------------------------------------------------

\newpage

%%%%%%%%%%%%%%%%%%%%%%%%%%%%%%%%%%%%%%%%%%%%
%%%%%%%%%%%%% FIGURE ( PLOTS )%%%%%%%%%%%%%
%%%%%%%%%%%%%%%%%%%%%%%%%%%%%%%%%%%%%%%%%%%%
%%%%%%%%%%%%%%%%%%%%%%%%%%%%%%%%%%%%%%%%%%%%%%%%%%%%%%%%%
%%%%%%%%%%%%%%%%%%%%%%%%%%%%%%%%%%%%%%%%%%%%%%%%%%%%%%%%%%
%%%%%%%%%%%%%%%%%%%%%%%%%%%%%%%%%%%%%%%%%%%%%%%%%%%%%%%%%
%%%%%%%%%%%%%%%%%%%%%%%%%%%%%%%%%%%%%%%%%%%%%%%%%%%%%%%%%
%%%%%%%%%%%%%%%%%%%%%%%%%%%%%%%%%%%%%%%%%%%%%%%%%%%%%%%%%
%%%%%%%%%%%%%%%%%%%%%%%%%%%%%%%%%%%%%%%%%%%%%%%%%%%%%%%%%
%%%%%%%%%%%%%%%%%%%%%%%%%%%%%%%%%%%%%%%%%%%%%%%%%%%%%%%%%
%%%%%%%%%%%%%%%%%%%%%%%%%%%%%%%%%%%%%%%%%%%%%%%%%%%%%%%%%

\end{document}